\documentclass{ws-procs9x6}

\begin{document}

\title{Unconventional Fermions: The price
of quark-lepton unification at TeV scales
\footnote{\uppercase{T}his work is supported in parts by the 
\uppercase{US} \uppercase{D}epartment of \uppercase{E}nergy
under \uppercase{G}rant \uppercase{N}o. \uppercase{DE-A505-89ER40518}.}}

\author{P.~Q.~HUNG}

\address{Dept. of Physics, University of Virginia, \\ 
382 McCormick Road\\
P. O. Box 400714, Charlottesville, Virginia 22904-4714, USA \\ 
E-mail: pqh@virginia.edu}

\maketitle

\abstracts{The early petite unification (PUT) of quarks and leptons at TeV
scales with $\sin^{2}\theta_{W}(M_{Z}^{2})$ used as a constraint,
necessitates the introduction of extra quarks and leptons with
unconventional electric charges (up to $4/3$ for the quarks
and $2$ for the leptons). This talk, in honor
of Paul Frampton's 60th birthday, will be devoted to the motivation and
construction of models of early unification and to their implications,
including the issues of rare decays and unconventional fermions.}

{\em Happy Birthday, Paul!}
\section{Motivations}

It is now believed that the Strong and Electroweak forces are very 
well described by $SU(3)_c \otimes SU(2)_L \otimes U(1)_Y$
at energies near or below the electroweak scale.
It is also widely believed that the SM is just a {\em low energy
manifestation} of some deeper unified theory which could
explain why the three gauge couplings are so different, why
the quantum numbers of the quarks are different from those of
the leptons, and (wishfully thinking) why fermion masses
are the way they are. Some of the
most successful and popular unification scenarios are the
quintessential 
$SU(5)$ or $SO(10)$ supplemented by supersymmetry.
With 3 couplings $\alpha_{3}(M_{Z}^{2})$,
$\alpha_{2}(M_{Z}^{2})$, $\alpha^{\prime}(M_{Z}^{2})$, one can make
two predictions: $M_{GUT} \approx 10^{16}\, GeV$
and $\sin^{2}\theta_{W}(M_{Z}^{2}) = 0.233(2)$.

Are there {\em alternatives} to GUT that can make predictions
for $\sin^{2}\theta_{W}(M_{Z}^{2})$ and that can be tested?
In particular, can these alternatives be constructed for energy scales
in the TeV region instead of being close to the Planck scale?
These are well motivated questions which are enhanced by recent interests
in the possibility that the ``fundamental scale'' lies in the TeV
region, in the context of large extra dimensions.
The key quantity used in the search for such alternatives is
\begin{equation}
\label{sin2}
\sin^{2}\theta_{W}(M_{Z}^{2}) = 0.23113(15) \,, 
\end{equation}
which is very precisely measured.

{\em Twenty two years ago}, a construction of an alternative
to GUT was made by Hung, Buras, and Bjorken \cite{HBB}
based on the data available at the time,
namely $\sin^{2}\theta_{W}(M_{Z}^{2}) \sim 0.22$.
The unification scale was found to be $\sim 1000\, TeV$, ``small''
enough to be coined the name {\em Petite Unification} (PUT) .

In light of the new and more precise data and of new
theoretical motivations, a reexamination of
PUT was performed by two of us (AB and PQH) \cite{BH}
yielding {\em three} possible scenarios
with some containing {\em unconventional fermions}. These three scenarios
predict the PUT scale to be less than 10 TeV. What is most
remarkable about two of the three scenarios is the existence of
these unconventional fermions which provide a {\em natural} way for
avoiding the severe constraint coming from the process
$K_L \rightarrow \mu e$, as we shall see below.

Since the concept of large extra dimensions involves scales in the
TeV range, it was natural to investigate the possibility of early
unification within the LED context. This, in fact, has been done
by Chacko, Hall, and Perelstein \cite{CHP} and by Dimopoulos and
Kaplan \cite{DK}. The model used in \cite{CHP} was, in fact, one of the
scenarios studied in \cite{HBB} transported to five dimensions. 

\section{A petite review of Petite Unification}

\subsection{What is Petite Unification?}

In any unification scheme, one would like to know what the predictions
might be. For example, in GUT, starting from {\em three fundamental couplings}:
$g_3,\,g_2,\,g^{\prime}$, one obtains  {\em one fundamental coupling}:
$g_{GUT}$, which results in two predictions: 
$M_{GUT}$ and $\sin^{2}\theta_{W}(M_{Z}^{2})$.

For Petite Unification, starting from {\em three fundamental couplings}:
$g_3,\,g_2,\,g^{\prime}$, one obtains  {\em two fundamental coupling}:
$g_S,\,g_W$, which results in one prediction: 
$\sin^{2}\theta_{W}(M_{Z}^{2})$ when the scale of Petite Unification
is constrained independently. The requirement that the scale is less than 10
TeV, for example, severely constrains the PUT gauge groups as we
shall see below.

We assume the PUT gauge group to be $G= G_S \otimes G_W$ with
the following pattern of symmetry breaking:
\begin{equation}
\label{pattern}
G \stackrel{\textstyle M}{\longrightarrow} G_1 
\stackrel{\textstyle \tilde{M}}{\longrightarrow} G_2
\stackrel{\textstyle M_Z}{\longrightarrow} SU(3)_c \otimes U(1)_{EM} ,
\end{equation}
where
\begin{equation}
\label{G1}
G_1 = SU(3)_{c}(g_3) \otimes \tilde{G}_{S}(\tilde{g}_S) \otimes
G_{W}(g_W) \, ,
\end {equation}
and
\begin{equation}
\label{G2}
G_2 = SU(3)_{c}(g_3) \otimes SU(2)_{L}(g_2) \otimes U(1)_{Y}(g^\prime)\,.
\end {equation}
It turns out that the most economical choices for $G_S$ and $G_W$ are the
Pati-Salam $SU(4)_{PS}$ \cite{PS} and $SU(N)^k$ respectively. The
PUT group is now
\begin{equation}
G= SU(4)_{PS} \otimes SU(N)^k \,, 
\end{equation}
with a permutation symmetry assumed so that 
each $SU(N)$ of $SU(N)^k$ has the same gauge coupling.

The next task is to compute $\sin^{2}\theta_{W}(M_{Z}^{2})$. Using
$Q = T_{3L} + T_{0}$ and $T_{0} = \sum_{\alpha} C_{\alpha W} 
T^{0}_{\alpha W} + C_{S} T_{15}$, and the matching of the
electromagnetic coupling with the weak couplings at $M_Z$, one
arrives at the following master formula for 
$\sin^{2}\theta_{W}(M_{Z}^{2})$ \cite{HBB}:
\begin{eqnarray}
\label{sinsq2}
\sin^{2}\theta_{W}(M_{Z}^{2})& =& \sin^{2}\theta_{W}^{0}[1-
C_{S}^{2}\frac{\alpha(M_{Z}^{2})}{\alpha_{S}(M_{Z}^{2})}-8\pi\times 
\nonumber \\
& &  \alpha(M_{Z}^{2})(K\ln\frac{\tilde{M}}{M_Z}+
K^{'}\ln\frac{{M}}{\tilde M})]\, ,
\end{eqnarray}
where
$\alpha(M_{Z}^{2}) \equiv e^{2}(M_{Z}^{2})/4\pi, \qquad
\alpha_{S}(M_{Z}^{2}) \equiv g_{3}^{2}(M_{Z}^{2})/4\pi$,
and
\begin{equation}
\label{sinsq0}
\sin^{2}\theta_{W}^{0} = \frac{1}{1+C_{W}^2} \,,
\end{equation}
with $C_{W}^2 = \sum_{\alpha}C_{\alpha W}^2$. Here 
\begin{equation}
\label{K}
K = b_1 - C_{W}^{2}b_2 - C_{S}^{2} b_3 \, ,
\end{equation}
\begin{equation}
\label{K'}
K^{'} = C_{S}^2 (\tilde{b} - \tilde b_3) \, .
\end{equation}
$\tilde{b}$ and $\tilde{b}_3$ are the one-loop Renomalization Group 
coefficients, above $\tilde{M}$, of $U(1)_S$ and $SU(3)_c$ respectively.
Furthermore, the following (fairly) precise inputs are used in
Eq. \ref{sinsq2}: $1/\alpha(M^2_Z)= 127.934(27), \,
\alpha_S(M^2_Z)= 0.1172(20)$.

From Eq. \ref{sinsq2}, it is important to realize the following point:
If we require $M$ and $\tilde{M}$ to be at most 10 TeV, the logarithmic
evolution of $\sin^{2}\theta_{W}$ is less important than in the GUT case.
In our case, a term which is crucial in the determination of 
$\sin^{2}\theta_{W}$ is the following term in Eq. \ref{sinsq2}:
$C_{S}^{2}\frac{\alpha(M_{Z}^{2})}{\alpha_{S}(M_{Z}^{2})}$. This can
easily be seen by looking at 
$\sin^{2}\theta_{W}(M_{Z}^{2})= 
\sin^{2}\theta_{W}^{0}(1-0.067\,C_{S}^2-\log terms)$. Indeed,
$\sin^{2}\theta_{W}(M_{Z}^{2})$ is found to be very sensitive to
the value of $C_{S}^2$ which, as we shall see, depends 
crucially on the fermion representation under $G_W$. Since we
have already chose $G_S$ to be $SU(4)_{PS}$, the choices of $G_W$
amount to their predictions for $\sin^{2}\theta_{W}^{0}$ and the
choices of fermion representations as represented by $C_{S}^2$.

Our objectives are therefore the computations of $\sin^{2}\theta_{W}^{0}$
and $C_{S}^2$, and to examine the related physical consequences.

\subsection{$\sin^{2}\theta_{W}^{0}$ and $C_{S}^2$}

The computation of $\sin^{2}\theta_{W}^{0}$ was done in detail in
\cite{HBB} and repeated in \cite{BH}. Here, I will simply state
the results. 

By the definition of $\sin^{2}\theta_{W}^{0}$ as shown
in Eq. \ref{sinsq0}, its computation requires simply the adjoint
representation of $G_W$. \cite{HBB} arrived at the following
important constraint: Only weak gauge bosons (i.e. the gauge bosons
of $G_W$) with charges $0,\pm1$ are consistent with the data.
We obtained a {\em very simple} formula for $\sin^{2}\theta_{W}^{0}$:
\begin{equation}
\label{simple2}
\sin^{2}\theta_{W}^{0} =\frac{N}{k Tr(Q_{W}^2)|_{adj}}= \frac{N}{k n_1}=
\frac{N}{2k r_0 (N-r_0)}, 
\end{equation}
with $n_1$ is the number of weak gauge bosons with charges $\pm 1$ and
$n_1=2 r_0 r_1$, with
$[\underbrace{\tilde{Q}_{W},\cdots \tilde{Q}_{W}}_{r_0},
\underbrace{\tilde{Q}_{W}-1,\cdots \tilde{Q}_{W}-1}_{r_1}]$.
The results for $\sin^{2}\theta_{W}^{0}$ are listed in Table~\ref{tab:table1} below.

\begin{table}
\tbl{The values of $\sin^2\theta^0$ for the 
weak groups $G_W=SU(N)^k$ and different fermion representations.\label{tab:table1}}
{\footnotesize
\begin{tabular}{ccccc} 
\hline
 & & &($f$,1)+(1,$\bar{f}$)& ($f$,$\bar{f}$) \\
$G_W$ & $r_0$ & $\sin^{2}\theta_{W}^{0}$ & $\tilde{Q}_{W}^{i}$ 
& $\tilde{Q}_{W}^{i}$ \\ \hline
$[SU(2)]^3$ & 1 & 0.333 & $\pm\frac{1}{2}$ & 0,$\pm1$ \\
$[SU(2)]^4$ & 1 & 0.250 & $\pm\frac{1}{2}$ & 0,$\pm1$ \\
$[SU(3)]^2$ & 1 & 0.375 & $\frac{2}{3}$,$-\frac{1}{3}$ & 0,$\pm1$ \\
$[SU(3)]^3$ & 1 & 0.250 & $\frac{2}{3},-\frac{1}{3}$ & 0,$\pm1$ \\
$[SU(4)]^2$ & 2 & 0.250 & $\pm\frac{1}{2}$ & 0,$\pm1$ \\
$[SU(5)]^2$ & 1 & 0.313 & $\frac{4}{5}$,$-\frac{1}{5}$ & 0,$\pm1$ \\
$[SU(6)]^2$ & 1 & 0.300 & $\frac{5}{6},-\frac{1}{6}$ & 0,$\pm1$ \\
$SU(7)$ & 3 & 0.292 & $\frac{4}{7},-\frac{3}{7}$&   \\
$[SU(7)]^2$ & 1 & 0.292 & $\frac{6}{7},-\frac{1}{7}$ & 0,$\pm1$ \\ 
$SU(8)$ & 3 & 0.267 & $\frac{5}{8},-\frac{3}{8}$&   \\
$SU(8)$ & 4 & 0.250 & $\pm\frac{1}{2}$&   \\
\hline
\end{tabular}}
\vspace*{-14pt}
\end{table}

Since $\tilde{Q}_{W}^{i} = \frac{1}{4}(3Q_{q}^{i} + Q_{l}^{i})$, one can see
that only groups and representations with $\tilde{Q}_{W}^{i} =
\pm\frac{1}{2}$ or $\tilde{Q}_{W}^{i} = 0, \pm 1$ can accomodate
standard fermions. With this in mind, Table 2 gives the
values of $C_{S}^2$ along with the corresponding quark and lepton charges.

\begin{table}
\tbl{The values of lepton ($Q^i_l$) and quark 
($Q^i_q$) electric charges  and the corresponding weak charge 
($\tilde Q^i_W$), and $C_{S}^2$.\label{tab:table2}}
{\footnotesize
\begin{tabular}{cccc}
\hline
$\tilde{Q}_{W}^{i}$& $Q_{l}^{i}$&$Q_{q}^{i}$&$C_{S}^2$ \\[1ex]
\hline
$\frac{1}{2}$& 0& $\frac{2}{3}$& \\[1ex]
$-\frac{1}{2}$&-1&$-\frac{1}{3}$& \\[1ex]
$\frac{1}{2}$&1&$\frac{1}{3}$&$\frac{2}{3}$ \\[1ex]
$-\frac{1}{2}$& 0& $-\frac{2}{3}$& \\[1ex] 
\hline
1&0&$\frac{4}{3}$& \\[1ex]
0&-1&$\frac{1}{3}$& \\[1ex]
-1&-2&$-\frac{2}{3}$&$\frac{8}{3}$ \\[1ex]
1&2&$\frac{2}{3}$& \\[1ex]
0&1&$-\frac{1}{3}$& \\[1ex]
-1&0&$-\frac{4}{3}$& \\[1ex] 
\hline
\end{tabular}}
\vspace*{-14pt}
\end{table}

From Table~\ref{tab:table1} and \ref{tab:table2}, and from
$\sin^{2}\theta_{W}(M_{Z}^{2})= 
\sin^{2}\theta_{W}^{0}(1-0.067\,C_{S}^2-\log terms)$, it is clear 
that groups with
``high'' $\sin^{2}\theta_{W}^{0}$ need ``high'' $C_{S}^2$.

A close examination revealed {\em three} favorite candidates
for $G_W$:

1. $[SU(2)]^4$: $C_{S}^{2}=2/3$; $\sin^{2}\theta_{W}^{0}=0.25$.

This group contains {\em only} conventionally-charged quarks
and leptons since the fermion representations under $G_W$ are 
of the type $(f,1,..)$ as one can easily infer from Tables 1 and 2. 

2. $[SU(2)]^3,\quad  [SU(3)]^2$: $C_{S}^{2}=8/3$;
$\sin^{2}\theta_{W}^{0}=1/3,\, 3/8$.

These groups contain conventionally-charged quarks and leptons
{\em as well as} unconventional quarks and leptons with higher
charges $\pm 4/3$ and $\pm 2$ as can be seen from Table 2. The
$G_W$-fermion representations are of the types: $(f,\bar{f},..)$.

The use of the term ``favorite'' actually means that these are the
three groups that can give $\sin^{2}\theta_{W}(M_{Z}^{2})$ within
the allowed experimental range for unification scales which are
less than 10 TeV. We shall see however that $[SU(2)]^4$ suffers
from problems with rare decays, and we will be left with 
$[SU(2)]^3,\quad  [SU(3)]^2$ as the true favorites.

In order to calculate $\sin^{2}\theta_{W}(M_{Z}^{2})$ or, equivalently,
the unification scale $M$, a knowledge of at least the fermionic degrees
of freedom that enter the evolution of $\sin^{2}\theta_{W}$ is necessary.

\section{Unconventional Fermions}

Although this section is titled ``unconventional fermions'', I will
list the fermion contents of all three ``favorite'' candidates.

I) $PUT_0=SU(4)_{\rm PS}\otimes 
SU(2)_{L} \otimes SU(2)_{R} \otimes \tilde{SU(2)}_{L} 
\otimes \tilde{SU(2)}_{R}$:

\begin{itemize}
\item Standard Fermions:
$\Psi_{L} = (q_L, l_L)=(4,2,1,1,1)_L$,
$\Psi_{R} = (q_R, l_R)=(4,1,2,1,1)_R$.
\item Fermions of ``Mirror Group'':
$\tilde{\Psi}_{L} = (\tilde{q}_L, \tilde{l}_L)
=(4,1,1,2,1)_L$, $\tilde{\Psi}_{R} = (\tilde{q}_R, \tilde{l}_R)
=(4,1,1,1,2)_R$.

\end{itemize}
It should be understood that the adjective ``Mirror'' refers to something
completely different (groups instead of fermions) 
from its customary use in Left-Right symmetric models.

II) $PUT_1=SU(4)_{\rm PS}\otimes 
SU(2)_{L} \otimes SU(2)_{H} \otimes {SU(2)}_{R}$:

Below the electric charges of the fermions are explicitely written down in
parentheses next to their names.

\begin{itemize}

\item Standard Fermions:

$\psi^{q}_{L,R} = \left(
\begin{array}{c}
u(2/3)\\ 
d(-1/3)
\end{array}
\right)_{L,R}$;
$\psi^{l}_{L,R} = \left(
\begin{array}{c}
\nu(0) \\ 
l(-1)
\end{array}
\right)_{L,R}$

\item  Fermions with ``weird'' charges:

$\tilde{Q}_{L,R} = \left(
\begin{array}{c}
\tilde{U}(4/3)\\ 
\tilde{D}(1/3)
\end{array}
\right)_{L,R}\ \,;\, \tilde{L}_{L,R} = \left(
\begin{array}{c}
\tilde{l}_{u}(-1)\\ 
\tilde{l}_{d}(-2)
\end{array}
\right)_{L,R}$

\item Fermion Representations:

$(4,2,2,1)_L =[(i\tau_{2} \psi^{q,*}_{L},\, \tilde{Q}_{L}), 
(\tilde{L}_{L},\, \psi^{l}_{L})]$;
$(4,1,2,2)_R =[(i\tau_{2} \psi^{q,*}_{R},\, \tilde{Q}_{R}), 
(\tilde{L}_{R},\, \psi^{l}_{R})]$;
$(4,2,1,1)_{L,R} = [\tilde{Q}^{\prime}_{L,R},\,
\tilde{L}^{\prime}_{L,R}]$;
$(4,1,1,2)_{L,R} = [\tilde{Q}^{\prime\prime}_{L,R},\,
\tilde{L}^{\prime\prime}_{L,R}]$


\item Tree-level $SU(2)_H$ transition:

$i\tau_{2} \psi^{q,*}_{L} \rightarrow 
\tilde{Q}_{L}$

$\tilde{L}_{L} \rightarrow \psi^{l}_{L}$

$\Rightarrow$ No tree-level transition 
between normal quarks and leptons due to $SU(2)_H$ gauge bosons

\item Tree-level $SU(4)/SU(3)_{c} \otimes
U(1)_S$ transition:

$i\tau_{2} \psi^{q,*}_{L} \rightarrow \tilde{L}_{L}$

$\tilde{Q}_{L} \rightarrow \psi^{l}_{L}$

etc...

$\Rightarrow$ No tree-level transition between normal
quarks and leptons due to PS gauge bosons.
\end{itemize}

III) $PUT_2=SU(4)_{\rm PS}\otimes 
SU(3)_{L} \otimes SU(3)_{H}$:

\begin{itemize}
\item Fermion Representations of the type $(4,3,\bar{3})$ and
$(4,\bar{3},3)$.

\item Same presence of higher charged fermions as in $PUT_1$!

\end{itemize}


Is the presence of quarks and leptons with unconventional charges
in $PUT_{1,2}$ a boon or a bane? As we shall see below, the existence
of these fermions turns into a virtue for $PUT_{1,2}$ 
when we look at the decay process $K_L\to \mu^\pm e^\mp$.
Before discussing the virtues and defects of these three scenarios, let
us do some RG analysis to see the range of values that the unification
scales can take.

\section{RG analysis and PUT scales}

\begin{itemize}
\item If $M$ and 
$\tilde{M}$ are of O(TeV), there is
not much ``running'' to do starting from $M_Z$. 
This means that two-loop contributions to 
$\sin^{2}\theta_{W}(M_{Z}^{2})$ are
not as important as the one-loop contribution. A detailed RG analysis
up to two loops within the context of $PUT_1$ will be presented in
\cite{BH2}. Here I will neglect that contribution for simplicity as
we had done in \cite{BH}.

\item To find PUT scales from $\sin^{2}\theta_{W}(M_{Z}^{2})$ in our 
RG analysis, we assume the unconventional fermions to have a mass
$M_F=(250\pm 50)\, GeV$. Furthermore,
we assume all vector-like fermions (present in 
$PUT_{1,2}$) to have a mass of order $M$.

\item Let us start out with $\tilde{M}=M$. We then use
Eq. (\ref{sin2}), namely $\sin^{2}\theta_{W}(M_{Z}^{2}) = 0.23113(15)$,
as a constraint to obtain $M$. Since the scale 
of $SU(2)_{R} \otimes \tilde{SU(2)}_{L} 
\otimes \tilde{SU(2)}_{R}$ breaking is naturally
of order $M$, we require (from the constraint on
$W_R$) that $M \geq 800 \, GeV$. Furthermore, we also require
$M_{F} \geq 200 \, GeV$.
We obtained the following results shown in Figure 1.




\begin{figure}[ht]
\centerline{\epsfxsize=3.1in\epsfbox{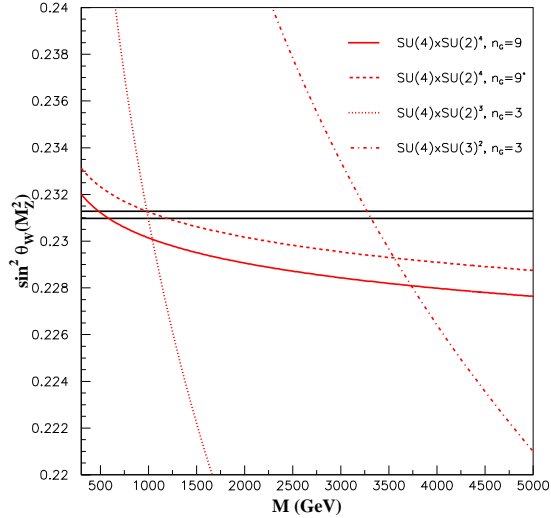}}   
\caption{$\sin^{2}\theta_{W}(M_{Z}^{2})$ versus the PUT scale
$M$. The horizonal band represents the experimental value.
The dashed curve ($n_G = 9^{*}$) is obtained by using $M_F = 200 GeV$,
while the other three curves are obtained by using $M_F = 250 GeV$.\label{fig1}}
\end{figure}







\item One can also look at the case where $\tilde{M} \neq M$. This
is shown in Figure 2 below.
\begin{figure}[ht]
\centerline{\epsfxsize=3.1in\epsfbox{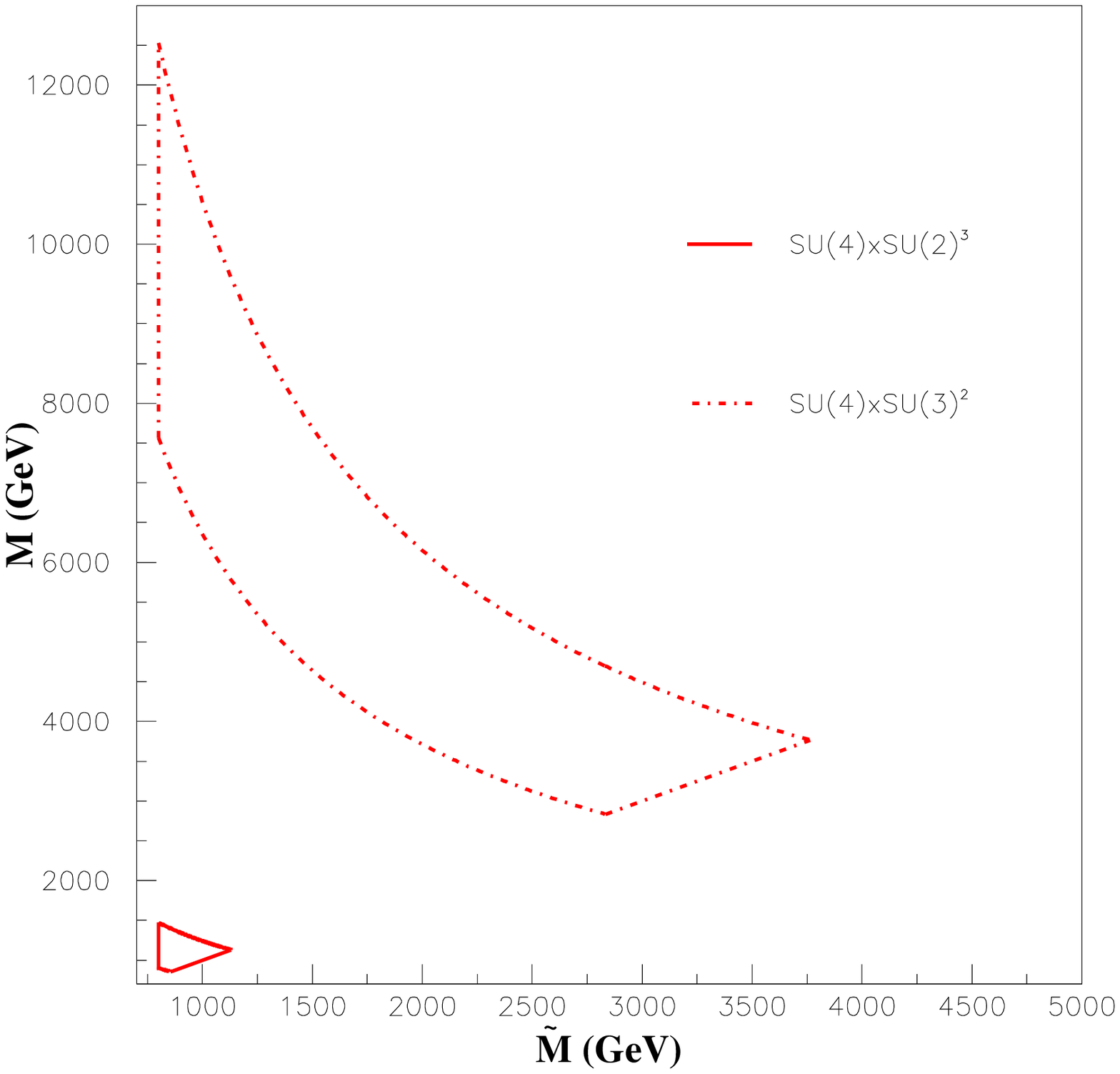}}   
\caption{The allowed ranges (at 2 $\sigma$) for the $SU(2)^3$ and 
$SU(3)^2$ scenarios.\label{fig2}}
\end{figure}

\end{itemize}

The following conclusions arise by examining Figure 1. 1) For $PUT_0$,
one needs the number of generations $n_{G} \geq 9$ for the SM and
$n_{G} \geq 4$ for MSSM. 2) For $PUT_1$ with $n_G=3$, one obtains
$M=(1.00\pm 0.14)$. 3) For $PUT_2$ with $n_G=3$, one obtains 
$M=(3.30 \pm 0.47)\, TeV$. 

From the above analysis, one can see that the PUT scales are all
below 10 TeV as promised.

\section{Virtues and Defects}

I will now discuss in particular the defects of $PUT_0$ and the
virtues of the unconventional fermions in $PUT_{1,2}$.


- $PUT_0$:

The defects are the following:


1) Large number of generations! (On the other hand, why not?)

2) Tree-level transition between
SM fermions which leads to a large $Br(K_L \rightarrow \mu e)$. 

$Br(K_L\to \mu^\pm e^\mp)= 4.7\cdot 10^{-12}
\left(\frac{\alpha_S(m_G)}{0.1}\right)^2 
 \times \left[\frac{1.8\cdot 10^3\,TeV}{m_G}\right]^4$

versus

$Br(K_L \rightarrow \mu e) < 
4.7 \times 10^{-12}$

Since $m_G \sim M < 1 \, TeV$, the bound is violated
by at least 13 orders of magnitude!

Chacko, Hall and Perelstein \cite{CHP} solved this problem by 
taking $PUT_0$ into five dimensions.

- $PUT_1$ ($PUT_2$):

The virtues of the unconventional fermions are as follows.

1) Tree-level transitions via $SU(2)_H$
and $SU(4)/SU(3)_{c} \otimes U(1)_S$ 
gauge bosons only occur between {\em unconventional} and {\em normal} fermions.
There is {\em no} tree-level FCNC.

The process $K_L \rightarrow \mu e$ occurs in box diagrams and
can be made small! 
(Exactly zero
when the unconventional fermions in the boxes are made degenerate.)

2) The lightest of the unconventional fermions (quark or lepton) is unstable.
It can decay entirely into normal fermions since the Higgs sector for the model
can mix $W_L,\, W_R,\,W_H$. For example, if $\tilde{l}_{u}(-1)$ were
the lightest of such particles (still presumably
having a mass larger than $M_W$), it can have the decay mode 
$\tilde{l}_{u}(-1) \rightarrow \nu W$. The rate will depend 
on the details of the mixing of the gauge bosons \cite{BH2}.

3) Since the lightest one is unstable, there is no cosmological constraint.

4) Fermions such as these unconventional ones can be searched 
for at the LHC (see \cite{FHS}),
especially if they are
relatively 'long lived''.

\section{Conclusions}

From our analysis, we have arrived at two favorite models:$PUT_{1}$ and
$PUT_{2}$. (A detailed study of $PUT_1$ will appear in the very near
future \cite{BH2}.)
The correct $\sin^{2}\theta_{W}(M_{Z}^{2})$ was obtained
for a PUT scale from 1-10 TeV.
These models predict an absence of tree-level FCNC because of the
presence of unconventionally charged quarks and leptons: The tree-level
transitions only connect these fermions to the normal ones!
Heavy (less than 1 TeV) and perhaps ''long lived''
quarkonic or leptonic unconventional fermions are characteristic 
signatures of this model.
Is this a heavy price to pay for early unification or an actual bonus?
Last but not least, since the unification scale is in the low TeV
region, one might wonder if there is any link to the physics of 
large extra dimensions.

\section*{Acknowledgments}
I wish to thank Paul Frampton and the organizers for a wonderful conference. 
Our thoughts go to Behram Kursunoglu who tirelessly organized the Coral Gables
series until he ultimately left us in October 2003.

\end{document}